%% file: main.tex
\documentclass[12pt]{article}

\usepackage[margin=1in]{geometry}
\usepackage{amsmath}
\usepackage{booktabs}
\usepackage{multirow}
\usepackage{graphicx}
\usepackage{threeparttable}
\usepackage{setspace}
\usepackage[round]{natbib}
\usepackage{microtype}
\usepackage[labelfont=bf, font=small, margin=0.5cm]{caption}
\usepackage{fancyhdr}
\usepackage[section]{placeins}

\usepackage[hidelinks]{hyperref}
\usepackage{tikz}
\usetikzlibrary{arrows.meta, positioning, shapes.geometric, fit, backgrounds}

\fancypagestyle{plain}{\fancyhf{}\fancyfoot[C]{\footnotesize\thepage}%
  }

\graphicspath{{figures/}}

\newcommand{\safeinput}[2]{\IfFileExists{#1}{\input{#1}}{\emph{[pending: #2]}}}

\title{Whose doctor does the AI recommend? An algorithm audit of reputation
and demographic signals in large language model--assisted physician choice}

\author{Syeda Anshrah Gillani$^{1}$ \qquad Mirza Samad Ahmed Baig$^{2}$\\[8pt]
\normalsize $^{1}$Heidelberg University, Heidelberg, Germany\\
\normalsize $^{2}$Fandaqah, Al Khobar, Saudi Arabia\\[6pt]
\normalsize Corresponding author: Syeda Anshrah Gillani\\
\normalsize \texttt{syeda.gilani@stud.uni-heidelberg.de}}
\date{}

\begin{document}

\maketitle

\input{sections/abstract.tex}

\section{Introduction}
\input{sections/intro.tex}

\section{Background and hypotheses}
\input{sections/background.tex}

\section{Methods}
\input{sections/methods.tex}

\section{Results}
\input{sections/results_scaffold.tex}

\section{Discussion}
\input{sections/discussion_skeleton.tex}

\bibliographystyle{plainnat}
\bibliography{refs}

\end{document}

%% file: sections/abstract.tex
\begin{abstract}
\noindent\textbf{Background:} Patients increasingly ask large language model
(LLM) assistants which doctor to see. These systems have become \emph{AI
infomediaries}: algorithms that intermediate a person's choice among other
people and thereby decide, silently and at scale, which physicians become
visible. What causally moves their recommendations, and whether physician
names signaling gender or ethnicity play a role, is undocumented, because
existing audits observe correlated real-world profiles from which causal
weights cannot be identified.

\noindent\textbf{Objective:} To estimate the causal effect of reputation
signals (patient rating, review volume and recency, feedback response,
hospital affiliation, visit fee, telehealth availability, years in practice),
name-signaled demographics, and list position on the probability that an LLM
recommends a physician.

\noindent\textbf{Methods:} Prespecified algorithm audit using a randomized
choice-based conjoint. Across three patient personas and nine prompt
paraphrases, each audited model chose among five synthetic family-medicine
physician cards whose attributes were independently randomized (3{,}024
choice sets; nine experimental arms probing decoding temperature, response
format, and presentation order). Physician gender and ethnicity were signaled
through names following correspondence-audit methodology. Average marginal
component effects (AMCEs) were estimated by linear probability models with
choice-set--clustered standard errors; effects are also expressed as
visit-fee equivalents.

\noindent\textbf{Results:} Across seven models (six open-weight;
gpt-4o-mini) and 40{,}068 scored responses, reputation signals dominate:
moving a rating from 3.9 to 4.7 raises choice probability by 31.4 pp (95\%
CI 30.5--32.3) and a \$90$\to$\$190 fee lowers it by 20.0 pp (CI
19.1--21.0). Demographic parity is rejected in an unexpected direction:
female-signaled names gain 2.5 pp (CI 1.8--3.2) and Hispanic-,
South-Asian--, and Black-signaled names gain 1.3--2.9 pp over
White-signaled names, tilts worth \$7--\$14 per visit in fee-equivalent
terms; being listed first is worth \$11. Yet models mentioned gender or
ethnicity in $\le$0.03\% of their stated reasons and abstained in 0.39\% of
trials: the demographic and position effects were invisible in the models'
own explanations. A reasoning model (deepseek-r1:7b) failed the
prespecified auditability gate outright.

\noindent\textbf{Conclusions:} AI infomediaries neither reproduce the
anti-minority discrimination documented in human audit studies nor achieve
neutrality: they silently apply systematic demographic tilts of their own.
The divergence between revealed weights and stated reasons means
transparency obligations that rely on model self-report would not surface
these effects; only behavioral audits can. The
frozen design makes the audit repeatable: any new model can be assessed
against identical stimuli with a single command, converting this study from
a one-time snapshot into a recurring monitoring instrument for AI-mediated
access to care.
\end{abstract}

\medskip
\noindent\textbf{Keywords:} large language models; algorithm audit;
algorithmic fairness; health equity; physician ratings; conjoint analysis;
patient choice; algorithmic gatekeeping; AI transparency; intersectional
bias; digital determinants of health; correspondence audit

%% file: sections/intro.tex
Choosing a doctor has always run on reputation. Patients weigh word of mouth,
online ratings, credentials, and cost, and a large literature documents how
these signals shape provider choice \citep{hanauer2014, emmert2013,
yaraghi2018}. A new intermediary has now inserted itself into that decision:
conversational assistants built on large language models (LLMs). Instead of
scanning a directory of physician profiles, a patient can simply ask an
assistant, ``Which of these doctors should I book?'' A growing share do, as
LLM assistants absorb search, comparison, and triage tasks that
previously belonged to rating platforms and search engines
\citep{ayers2023}. When an assistant answers, it collapses a multi-attribute
comparison into a single recommendation. The weights it applies, whether to
a star rating, a consultation fee, a hospital affiliation, or, more
troublingly, to a physician's name, take effect silently, at scale, and
without any of the professional accountability that governs human referral.
The assistant has become an \emph{AI infomediary}: an algorithm that
intermediates one person's choice among other people, and in doing so
allocates visibility, and ultimately patients, across physicians.

This paper asks a simple causal question: \emph{what moves an LLM's physician
recommendation?} We answer it with a prespecified algorithm audit
\citep{sandvig2014, metaxa2021} built on a randomized choice-based conjoint
design \citep{hainmueller2014}. Audited models repeatedly choose among five
synthetic family-medicine physician cards whose attributes are independently
randomized: patient rating, review volume, review recency, whether the
practice responds to feedback, hospital-system affiliation, new-patient visit
fee, telehealth availability, years in practice, and gender and ethnicity,
which are signaled through physician names using correspondence-audit
methodology \citep{bertrand2004, gaddis2017}. Because every attribute varies
independently of every other and of display order, differences in choice
probability identify the average marginal component effect (AMCE) of each
signal, and the fee attribute converts any effect into an interpretable
dollars-per-visit equivalent.

The design speaks to three literatures at once. First, it extends the
economics of online physician reputation \citep{gao2012, yaraghi2018} from
human choosers to algorithmic ones: prior conjoint evidence shows how
\emph{patients} trade off ratings and other attributes; we provide the
matching causal estimates for the AI assistants now advising them. Second, it
extends the algorithmic-fairness literature in health care
\citep{obermeyer2019, zack2024, omiye2023}, which has concentrated on
clinical decisions (triage, diagnosis, treatment), to the consumer-facing
question of \emph{which humans the algorithm makes visible}. Existing audits
of chatbot specialist recommendations are descriptive, asking models to name
real physicians and characterizing who appears \citep{oca2023}; because real
physicians' attributes are correlated, such designs cannot separate a rating
effect from a name effect. Our synthetic-profile randomization can. Third, it
contributes to the emerging study of AI infomediaries and generative engine
optimization \citep{aggarwal2024}, where a companion audit in the travel
domain \citep{baig2026} found that assistants reproduce human valence-price
primacy while introducing content-free position effects; identical rating and
recency levels here permit the first cross-domain comparison of LLM
signal weights.

Concretely, we make four contributions. (1) We provide, to our knowledge, the
first randomized conjoint audit of LLM physician recommendation, across a
panel of widely deployed models, three patient personas, nine prompt
paraphrases, and nine prespecified experimental arms. (2) We estimate
fee-equivalent values for every signal, including the dollar value of a
first-listed position and of name-signaled demographic attributes, with
uncertainty quantified by cluster-robust inference and Krinsky--Robb
simulation. (3) We test demographic parity directly, at the group level and
at the intersection of gender and ethnicity, using prespecified equivalence
bounds, so that a null is informative rather than merely underpowered, and
we characterize when models abstain from choosing or spontaneously flag
demographic attributes. (4) We compare the models' \emph{stated} reasons against their
\emph{revealed} weights, quantifying how faithfully AI explanations track AI
behavior in a health context. Beyond these estimates, the
frozen-design instrument makes every finding falsifiable and repeatable:
auditing any future model against identical stimuli is a single command, so
the parity verdicts reported here can be re-issued, or revoked, with each
model release.

Across seven models and 40{,}068 scored responses, three findings emerge.
First, LLM recommenders run on reputation: patient rating alone carries
37.7\% of total attribute importance and fee 24.1\%, with a 3.9$\to$4.7
rating step worth \$157 per visit in fee-equivalent terms. Second,
demographic parity fails, but not in the direction human audit studies
would predict: female-signaled and minority-signaled names \emph{gain} one
to three percentage points, tilts worth \$7--\$14 per visit, and no
gender$\times$ethnicity cell meets the prespecified equivalence bound.
Third, none of this is visible in the models' own explanations: gender and
ethnicity appear in fewer than 0.03\% of stated reasons, and models
abstained from choosing in only 0.39\% of trials, so the demographic and
position effects operate entirely below the models' self-reported
rationale.

The remainder of the paper reviews the relevant literatures and derives
hypotheses (Section~2), details the design, model panel, and estimation
strategy (Section~3), reports results (Section~4), and discusses implications
for platforms, regulators, and health-services research (Section~5).

%% file: sections/background.tex
\subsection{Online reputation and physician choice}
Physician-rating websites have made service reputation legible in health
care. Public awareness and use of these ratings grew rapidly through the
2010s \citep{hanauer2014}, physician coverage on rating platforms expanded in
parallel \citep{gao2012}, and systematic reviews document that valence and
volume of reviews influence patients' stated willingness to select a provider
\citep{emmert2013}. The causal benchmark most relevant to our design is
\citet{yaraghi2018}, a choice-based conjoint in which patients chose between
physician profiles with randomized quality ratings: higher ratings raised
choice odds substantially, and nonclinical (service) ratings mattered
alongside clinical ones. Beyond ratings, patients weigh access attributes
(wait time, fees, telehealth) and credential attributes (experience,
affiliation) in ways that vary with their situation: an uninsured patient
weighs fees differently from a chronically ill one. Our persona manipulation
mirrors that heterogeneity.

If LLM assistants have internalized the regularities of human choice
documented in this literature, as their training corpora suggest they
should, their revealed weights should reproduce the canonical ordering:
valence first, then price, with secondary signals (volume, recency,
responsiveness, affiliation, experience, telehealth) positive but smaller.
Hypotheses H1--H8 formalize this expectation attribute by attribute
(Table~\ref{tab:confirmatory} lists the full prespecified family):
higher patient ratings (H1), more reviews (H2), more recent reviews (H3),
practices that respond to feedback (H4), university-hospital affiliation
(H5), lower visit fees (H6), telehealth availability (H7), and more years in
practice (H8) each increase the probability of recommendation.

\subsection{Names as treatments: discrimination audits meet AI advice}
Correspondence audits established that names alone shift decisions: résumés
signed ``Emily'' or ``Greg'' received 50\% more callbacks than identical
résumés signed ``Lakisha'' or ``Jamal'' \citep{bertrand2004}, and the
methodology for selecting demographically distinctive names is now standard
\citep{gaddis2017}. Field experiments extended the finding to platform
markets \citep{edelman2017}. In health care specifically, algorithmic systems
have repeatedly encoded group disparities: a widely used care-management
algorithm systematically under-referred Black patients \citep{obermeyer2019},
and LLMs have been shown to propagate race-based medicine
\citep{omiye2023} and to generate clinical vignettes and treatment
recommendations that vary with patient demographics \citep{zack2024}.

That literature concerns the algorithm's treatment of \emph{patients}. The
question here is different and unexamined: when the algorithm chooses among
\emph{physicians}, does a name signaling gender or ethnicity move the
recommendation, holding every reputational attribute fixed? Three outcomes
are possible, and all are informative. The assistant may reproduce human
discrimination absorbed from training data; it may be neutral, because
alignment training suppresses demographic reasoning; or it may overcorrect.
We therefore prespecify parity nulls with equivalence bounds rather than
directional predictions: name-signaled gender (H9) and ethnicity (H10) have
no effect on recommendation probability, tested jointly across the panel and
bounded by two one-sided tests at $\pm$1.5 percentage points. Because
demographic effects in health care are often intersectional, with
clinical-bias audits reporting disparities for specific
gender$\times$ethnicity combinations that neither main effect anticipates
\citep{zack2024}, we additionally test
whether the two signals combine additively (H13) and estimate all ten
gender$\times$ethnicity cell effects. Finally, an abstention is not
necessarily a defect: when only names distinguish otherwise-comparable
physicians, a model that declines to choose behaves more defensibly than one
that silently lets names decide. We therefore treat abstention and whether
responses explicitly acknowledge demographic attributes as outcomes in their
own right.

\subsection{Position, personas, and explanation faithfulness}
Two further properties of LLM choice behavior carry over from algorithm
audits outside health. First, display position: retrieval and recommendation
interfaces reward rank, and our companion travel-domain audit
\citep{baig2026} found a causal first-listed advantage worth real money per
booking. Because position is content-free, any position effect in physician
recommendation is a pure artifact of the medium, consequential for
provider-directory platforms whose orderings feed AI assistants. Second,
context sensitivity: H11 predicts that the fee penalty is larger for a
persona paying out of pocket, and H12 that review volume moderates the rating
effect (a 4.7 average over 400 reviews is more diagnostic than over 12
\citep{yaraghi2018}). Finally, models explain their picks on request, but
stated reasons need not track revealed weights; in the travel audit they
tracked imperfectly \citep{baig2026}. We quantify that gap here, with
particular attention to whether demographic attributes that move choices are
ever \emph{mentioned} in explanations, the failure mode with the clearest
governance implications.

%% file: sections/methods.tex
\subsection{Design overview}
We conducted a randomized choice-based conjoint embedded in an algorithm
audit. Each trial presented one audited model with a persona, a prompt
template, and five synthetic physician cards, and asked it to recommend one.
Cards described board-certified family-medicine physicians accepting new
patients with a practice near the user (all held constant), while eight
reputational attributes varied independently and uniformly at random per
card: patient rating (3.9, 4.3, or 4.7 of 5), number of reviews (12, 85, or
400), most recent review (3 days or 11 months ago), whether the practice
responds to patient feedback (present or absent), affiliation (university
hospital system vs.\ independent practice), new-patient visit fee (\$90,
\$140, or \$190), telehealth availability (present or absent), and years in
practice (8, 18, or 28).
Rating and recency levels replicate our travel-domain audit \citep{baig2026}
to permit cross-domain comparison. Display order (slot 1--5) was randomized
independently of content, identifying list-position effects.
Figure~\ref{fig:pipeline} summarizes the pipeline from frozen design to
inference; Figure~\ref{fig:choiceset} shows one card verbatim.

\input{figures/fig_pipeline.tex}
\input{figures/fig_choiceset.tex}

Physician gender (female/male) and ethnicity (White, Black, Hispanic, East
Asian, South Asian) were randomized per card and signaled through names
(``Dr.\ Priya Sharma''), with four first names and four surnames per
gender$\times$ethnicity cell drawn from the correspondence-audit tradition
\citep{bertrand2004, gaddis2017}. Analysis is at the cell level; a
prespecified placebo tests that the specific name exemplar within a cell has
no effect. Within a choice set, no two cards shared a surname or a full
attribute profile.

\subsection{Stimuli, personas, and arms}
Prompts crossed three patient personas (new in town; managing a chronic
condition; uninsured and paying out of pocket) with nine paraphrased
templates, and requested a JSON response naming the chosen physician and a
one-to-two-sentence reason. The design comprises 3{,}024 main-arm choice sets
(112 per persona$\times$template stratum), fixed across all audited models:
every model sees identical stimuli, so model comparisons are within-trial.
Eight prespecified secondary arms re-present stratified subsets of the same
trials under altered conditions: temperature 0 (594 trials), grounded
web-snippet formatting (297), top-3 ranking (297), reason-before-choice field
order (189), five-repetition test--retest (108), a letter-labeled logprob arm
separating position from letter-token preference (297), a persona-free
control (297), and reversed card-line order (189). A 20-repetition
manipulation check asked each model to define board certification. The design
matrix was generated once (seed 20260702), hashed (SHA-256), and frozen
before any non-pilot model call, and the analysis plan (hypotheses,
estimators, equivalence bounds, and exclusion rules) was written and fixed
before any confirmatory data were analyzed.

\subsection{Model panel and execution}
The audited panel comprises six widely deployed open-weight
instruction-tuned models executed locally via Ollama (llama3.2:3b,
qwen2.5:3b, phi3:mini, mistral:7b-instruct-q4\_K\_M, gemma3:4b, and
llama3.1:8b) and one proprietary production model (gpt-4o-mini, accessed
through an Azure OpenAI deployment, API version 2025-01-01-preview) under
the identical protocol, including seed control and, on the letter arm,
token-level log probabilities; exact identifiers and inference dates are
recorded in the design metadata. A seventh open-weight candidate, deepseek-r1:7b, was
subject to a prespecified pilot gate and excluded before full-run
collection (Results). Generation used
temperature 0.7 (top-$p$ 0.9) except in the temperature arm, with
per-call deterministic seeds derived from (model, arm, trial, repetition).
Responses failing JSON parsing or naming no listed physician were retried up
to twice with a stricter reminder; remaining failures are analyzed as
missingness (prespecified exclusion of model$\times$arm cells exceeding 5\%
failure). All calls are checkpointed and the pipeline is resume-safe.

\subsection{Exploratory frontier-model pilot and archived forecasts}
Because API access to frontier models was not available at collection time, we
additionally piloted four Claude-family models (claude-haiku-4-5,
claude-sonnet-5, claude-opus-4-8, claude-fable-5) on the first 50 main-arm
choice sets of the frozen design. These responses were \emph{not} collected
under the audit protocol above: they were gathered interactively inside an
agentic coding environment (Claude Code, Anthropic), one fresh subagent per
trial with no shared context, using the pipeline's own stimulus renderer and
parser (verbatim frozen prompts; one stricter re-ask on unparseable output;
0/200 parse failures). This collection mode leaves decoding parameters at the
harness defaults rather than the protocol's temperature $0.7$/top-$p$ $0.9$,
embeds each response in an agent system prompt rather than a bare API call,
and is not seed-reproducible. The 200 resulting records are therefore
quarantined from the analysis pipeline, enter no confirmatory test, and are
reported only descriptively (Table~\ref{tab:pilot_frontier}).

We put this pilot to a second, falsifiable use. For each piloted model we fit
a conditional logit to its 50 choice sets and computed out-of-sample forecast
probabilities of being chosen for every card in all 3{,}024 main-arm choice
sets. These forecasts, which are model-based extrapolations rather than
observations, were
archived, with a timestamp, \emph{before} any full API run of these
models, so that a subsequent protocol-compliant run (claude-haiku-4-5 is the
designated target) can grade them: agreement would suggest small
harness-contaminated pilots cheaply approximate full audits; disagreement
would itself measure the harness and decoding effects described above.

\subsection{Estimation}
The estimand is the AMCE of each attribute level on the probability of
being recommended \citep{hainmueller2014}. Primary estimator: linear
probability regression of an indicator for being chosen on attribute-level
dummies and slot dummies, with standard errors clustered on choice set,
estimated per model and pooled with model fixed effects. Secondary:
conditional (McFadden) logit per choice set. Thirteen prespecified hypotheses
are tested by joint Wald tests with Holm correction; demographic parity
hypotheses (H9--H10) are additionally assessed by two one-sided tests with a
smallest-effect-size-of-interest of $\pm$1.5 percentage points, and are
pooled estimands, because a Monte-Carlo precision analysis showed per-model power
for plausible demographic effect sizes is inadequate, whereas the pooled
panel yields a minimum detectable effect below 1 percentage point. Because a
linear-probability interaction can be nonzero under a purely additive
logit process, H12 is corroborated only if the conditional-logit interaction
agrees in sign and significance. H13 tests intersectionality, that is,
whether name-signaled gender and ethnicity combine additively, as a joint test of
their interaction terms; we additionally report all ten gender$\times$ethnicity
cell effects relative to the White-male reference, each with an equivalence
verdict. Separately and descriptively, we classify each response as an
abstention (declining to choose) and as demographic-aware (referring to a
physician's name, gender, ethnicity, or race), the latter measuring how often
a model's own words reveal that a name-signaled attribute was salient. Fee-equivalents divide each AMCE by the
per-dollar fee AMCE (linearized over the \$100 range), with 95\% intervals by
Krinsky--Robb simulation (10{,}000 draws), gated on monotonicity of the fee
response. Stated reasons are coded for attribute mentions by a validated
dictionary and an independent judge model (gemma2:2b, excluded from the
audited panel), and compared with revealed importance shares.

\subsection{Validation and reproducibility}
The complete pipeline (stimulus rendering, execution, parsing, and every
analysis module) was validated end-to-end against a mock data-generating
process with known conditional-logit effects, including planted name-signaled
effects (female $+0.10$; Black $-0.15$ on the logit scale) and true-null
ethnicity effects. The validation suite (18 assertions) confirms recovery of
signs, magnitudes, true nulls, the fee-monotonicity gate, the name-exemplar
placebo, and figure/table generation. Every call is checkpointed, and the
frozen design matrices are hash-verified at load time, so a full audit is
reproducible from the design seed. All physician identities are
synthetic, and the demographic manipulation audits model behavior, not real
physicians.

%% file: figures/fig_pipeline.tex
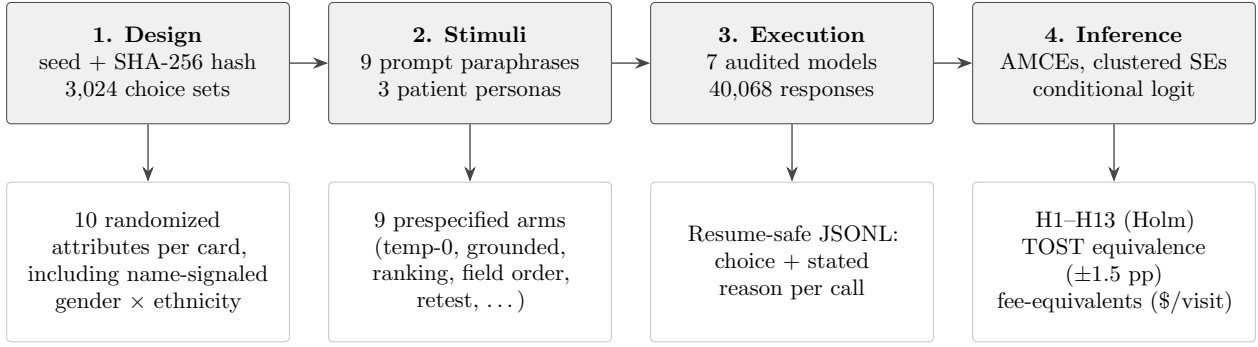
\begin{figure}[tbp]
\centering
\resizebox{\linewidth}{!}{%
\begin{tikzpicture}[
  font=\footnotesize,
  stage/.style={draw=gray!55!black, rounded corners=2pt, align=center,
                inner sep=6pt, text width=40mm, minimum height=19mm,
                fill=gray!12},
  detail/.style={draw=gray!45, rounded corners=2pt, align=center,
                 inner sep=6pt, text width=40mm, minimum height=25mm,
                 fill=white},
  arr/.style={-{Stealth[length=2.4mm]}, semithick, gray!55!black}
]

\node[stage]                  (s1) {\textbf{1. Design}\\
  seed $+$ SHA-256 hash\\ 3{,}024 choice sets};
\node[stage, right=6mm of s1] (s2) {\textbf{2. Stimuli}\\
  9 prompt paraphrases\\ 3 patient personas};
\node[stage, right=6mm of s2] (s3) {\textbf{3. Execution}\\
  7 audited models\\ 40{,}068 responses};
\node[stage, right=6mm of s3] (s4) {\textbf{4. Inference}\\
  AMCEs, clustered SEs\\ conditional logit};

\node[detail, below=9mm of s1] (d1) {10 randomized\\
  attributes per card,\\ including name-signaled\\
  gender $\times$ ethnicity};
\node[detail, below=9mm of s2] (d2) {9 prespecified arms\\
  (temp-0, grounded,\\ ranking, field order,\\ retest, \ldots)};
\node[detail, below=9mm of s3] (d3) {Resume-safe JSONL:\\
  choice $+$ stated\\ reason per call};
\node[detail, below=9mm of s4] (d4) {H1--H13 (Holm)\\
  TOST equivalence\\ ($\pm$1.5 pp)\\
  fee-equivalents (\$/visit)};

\draw[arr] (s1) -- (s2);
\draw[arr] (s2) -- (s3);
\draw[arr] (s3) -- (s4);

\draw[arr] (s1) -- (d1);
\draw[arr] (s2) -- (d2);
\draw[arr] (s3) -- (d3);
\draw[arr] (s4) -- (d4);

\end{tikzpicture}}
\caption{Audit pipeline. A design frozen (and hashed) before any data
collection generates 3{,}024 five-card choice sets whose attributes vary
independently; each audited model answers every choice set under nine
prespecified arms; average marginal component effects and prespecified
equivalence tests are estimated from the pooled choices. Because
randomization is at the attribute level, any systematic difference in
choice probability identifies the causal weight the model places on that
signal.}
\label{fig:pipeline}
\end{figure}

%% file: figures/fig_choiceset.tex
\begin{figure}[tbp]
\centering
\resizebox{\ifdim\width>\linewidth \linewidth\else \width\fi}{!}{%
\begin{tikzpicture}[font=\footnotesize]

\node[draw=gray!55!black, rounded corners=3pt, inner sep=9pt, align=left,
      fill=gray!4, font=\footnotesize\ttfamily] (card) {%
Dr.\ Priya Sharma \textcolor{gray!60!black}{- Family medicine physician}\\
\textcolor{gray!60!black}{Board-certified, accepting new patients}\\
\textcolor{gray!60!black}{Practice located within 20 minutes of you}\\
Patient rating: 4.7/5 (85 reviews)\\
Most recent review: 3 days ago\\
In practice for 18 years\\
New-patient visit fee: \$140\\
Affiliated with a major university hospital system\\
Practice responds to patient feedback\\
Telehealth visits available};

\node[anchor=north west, align=left, text width=55mm]
      at ([xshift=8mm]card.north east) {%
\hyphenpenalty=10000\exhyphenpenalty=10000\relax
\textbf{Randomized per card}\\[1mm]
name (gender $\times$ ethnicity signal), patient rating, review volume,
review recency, response to feedback, hospital affiliation, visit fee,
telehealth, years in practice\\[3.5mm]
\textcolor{gray!60!black}{\textbf{Held constant}}\\[1mm]
\textcolor{gray!60!black}{specialty, board certification,\\
distance, accepting new patients}};

\end{tikzpicture}}
\caption{One physician card as presented to the models (card layout shown;
the grounded-arm snippet and reversed-layout variants are described in
Methods). Black lines vary independently between cards; gray lines are
identical on every card, so specialty, board certification, and distance
cannot drive choices. Each choice set presents five such cards, and the
model is asked to choose one for the persona and give a brief reason.}
\label{fig:choiceset}
\end{figure}

%% file: sections/results_scaffold.tex
\subsection{Data quality and manipulation checks}
The confirmatory panel comprises seven models: six open-weight
instruction-tuned models (llama3.2:3b, qwen2.5:3b, phi3:mini,
mistral:7b-instruct, gemma3:4b, llama3.1:8b) and one proprietary production
model (gpt-4o-mini, accessed via Azure OpenAI), yielding 40{,}068 scored
responses across the nine arms. Parse-failure rates were 0\% for gpt-4o-mini
on every arm and below 2\% for nearly all model--arm cells
(Table~\ref{tab:parse}); the exceptions were mistral on the retest (32.0\%)
and top-3 ranking (14.1\%) arms, which enter the missingness analysis in
Section~4.7. An eighth candidate, deepseek-r1:7b, failed its prespecified
pilot gate (100\% parse failures: its chain-of-thought exhausted the fixed
output budget before emitting the required JSON) and was excluded before any
full-run data were collected. This is itself a finding about the auditability of
reasoning models under fixed-format protocols. Test--retest consistency was
high: the modal choice was repeated in 73\% (mistral) to 99\% (gemma3) of
five-fold repetitions (Table~\ref{tab:retest}).

\safeinput{tables/tab_parse.tex}{parse-failure table}
\safeinput{tables/tab_retest.tex}{retest table}

\subsection{What moves the recommendation: AMCEs}
\begin{figure}[t]
\centering
\includegraphics[width=\linewidth]{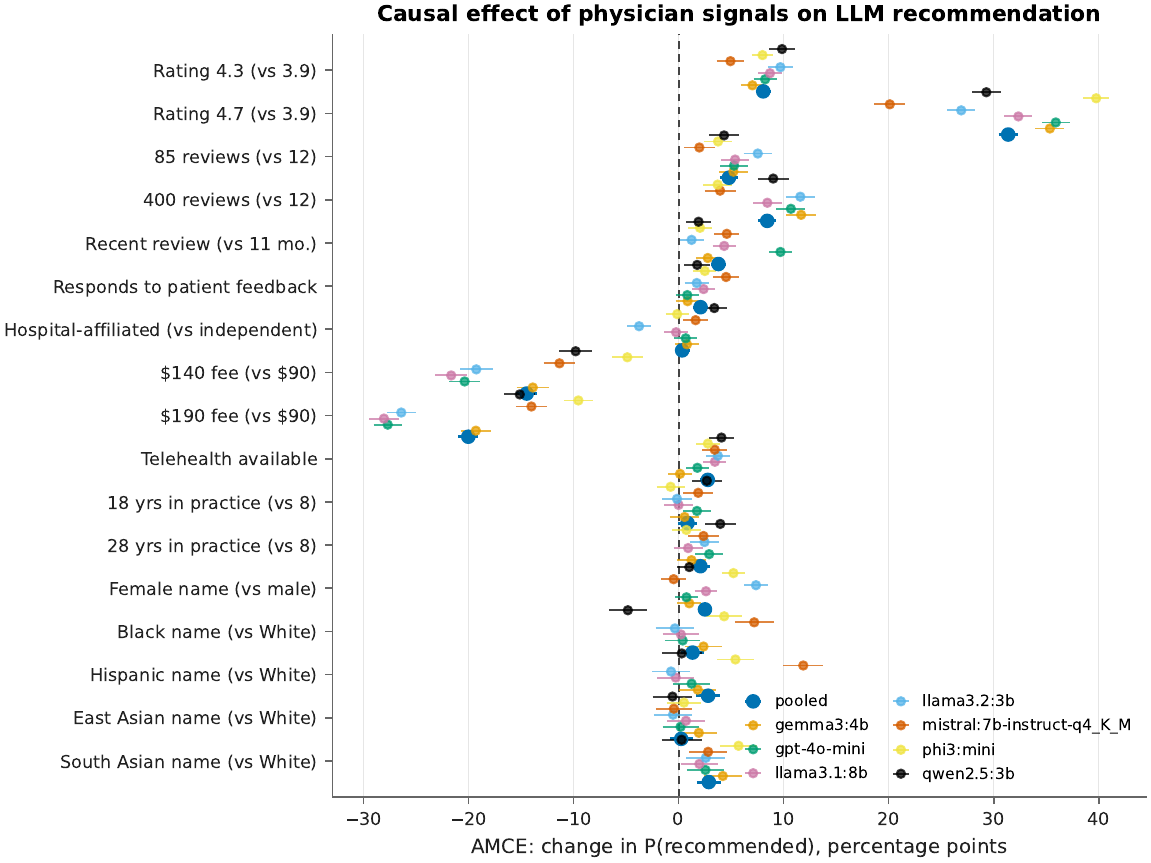}
\caption{Average marginal component effects on choice probability (pooled
and per model), with 95\% choice-set--clustered confidence intervals.
Baseline choice probability is 20\% (five cards per set).}
\label{fig:amce}
\end{figure}

Reputation signals dominate (Figure~\ref{fig:amce};
Table~\ref{tab:amce_main}). Pooled across
models, moving a card's rating from 3.9 to 4.7 raises its choice probability
by 31.4 percentage points (pp) against a 20\% baseline (95\% CI
30.5--32.3); raising the fee from \$90 to \$190 lowers it by 20.0 pp (CI
19.1--21.0). Review volume (400 vs.\ 12 reviews: $+8.4$ pp), review recency
($+3.8$ pp), telehealth availability ($+2.8$ pp), responding to feedback
($+2.1$ pp), and 28 vs.\ 8 years in practice ($+2.1$ pp) all move choices in
the expected direction. Hospital-system affiliation does not ($+0.3$ pp,
$P=.35$), the lone prespecified reputation hypothesis (H5) not supported.
In share-of-importance terms, rating accounts for 37.7\% of the total
attribute range, price 24.1\%, volume 10.1\%, and list position 8.2\%
(Table~\ref{tab:importance}). Confirmatory hypotheses H1--H4 and H6--H8 were
all supported after Holm correction (Table~\ref{tab:confirmatory}).

\safeinput{tables/tab_amce_main.tex}{main AMCE table}
\safeinput{tables/tab_confirmatory.tex}{confirmatory tests table}
\safeinput{tables/tab_importance.tex}{importance shares table}

\subsection{Name-signaled gender and ethnicity}
Demographic parity is rejected, in a direction the bias literature would
not predict. Physician cards bearing female-signaled names were chosen 2.5
pp more often than male-signaled ones (CI 1.8--3.2; H9 $P_{\text{Holm}}<.001$), and ethnicity effects are jointly significant (H10
$P_{\text{Holm}}<.001$): relative to White-signaled names,
Hispanic-signaled ($+2.8$ pp), South-Asian--signaled ($+2.9$ pp), and
Black-signaled ($+1.3$ pp) names were chosen \emph{more} often, with
East-Asian--signaled names indistinguishable from White ($+0.3$ pp,
$P=.65$). The conditional-logit specification corroborates the linear
estimates (pooled odds ratios 1.22 for female, 1.15--1.24 for the elevated
ethnicity cells). Because the prespecified TOST bounds of $\pm$1.5 pp are
excluded by several of these intervals, the correct summary is not
``parity holds'' but ``models exhibit a systematic \emph{pro-female,
pro-minority} tilt of one to three percentage points.'' The prespecified
name-exemplar placebo, however, is violated: choice shares differ across
name exemplars \emph{within} gender$\times$ethnicity cells (LR $=370.0$,
df $=30$, $P<.001$), so cell-level effects should be read as
perceived-category effects entangled with name-specific familiarity, and
the demographic estimates as audit-level signals rather than precise
category parameters.

\subsubsection{Intersectional effects}
The gender$\times$ethnicity interaction is not jointly significant after
correction (H13, $P_{\text{Holm}}=.099$), indicating approximately additive
effects, but the additive combination concentrates advantage: relative to
White-male cards, female-Hispanic cards gain 5.7 pp (CI 4.0--7.3),
female--South-Asian 5.1 pp, and female-Black 3.9 pp
(Table~\ref{tab:intersectional}). No cell meets the prespecified
equivalence criterion, so no gender$\times$ethnicity combination can be
certified as neutrally treated.

\begin{figure}[t]
\centering
\includegraphics[width=\linewidth]{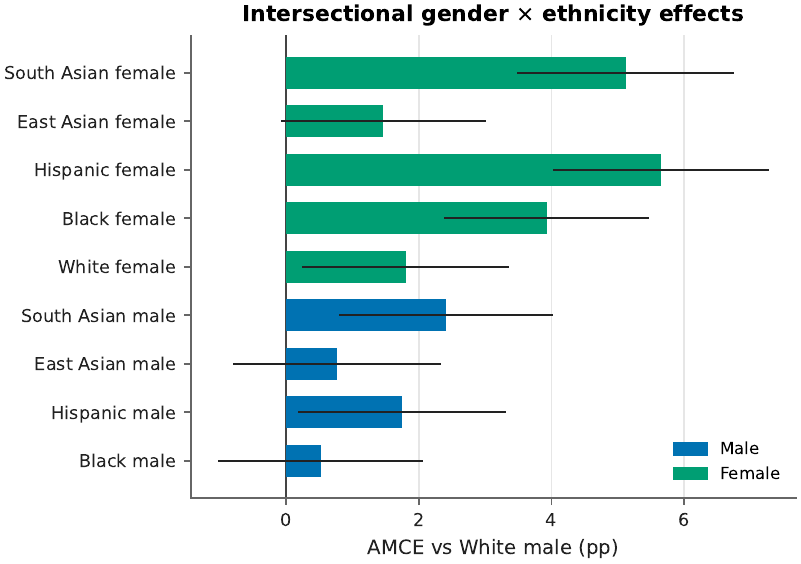}
\caption{Choice-probability advantage of each gender$\times$ethnicity cell
relative to White-male cards, with 95\% confidence intervals and the
prespecified $\pm$1.5 pp equivalence band.}
\label{fig:intersectional}
\end{figure}

\safeinput{tables/tab_intersectional.tex}{intersectional cell table}

\subsubsection{Abstention and demographic-aware responses}
Abstention is rare and indiscriminate: models declined to choose in 0.39\%
of responses overall, and spontaneously flagged demographic attributes in
only 0.01\% (Table~\ref{tab:refusals}). The models that let names move
their choices essentially never said so.

\input{tables/tab_refusals.tex}

\subsection{Position effects}
\begin{figure}[t]
\centering
\includegraphics[width=0.85\linewidth]{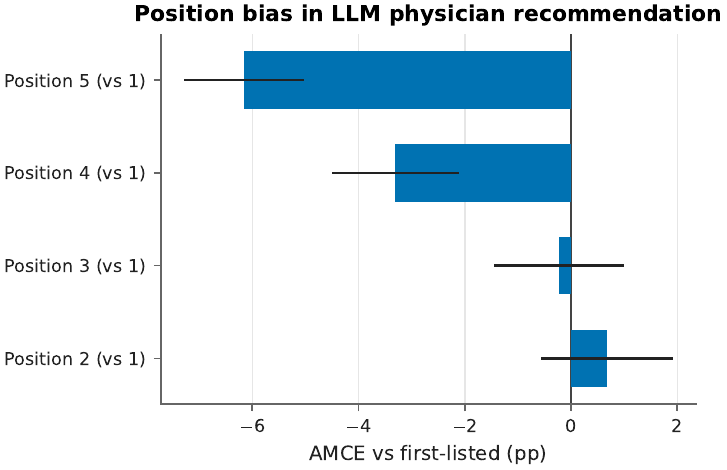}
\caption{List-position effects on choice probability by model, holding card
content constant. Slot 1 is the reference.}
\label{fig:position}
\end{figure}

Position effects are content-free and economically meaningful: cards in
slot 5 lose 6.2 pp (CI 5.0--7.3) and slot 4 loses 3.3 pp relative to slot
1, holding all content constant (Figure~\ref{fig:position}). The letter-labeled arm decomposes
label from position using gpt-4o-mini token logprobs: both components are
small in isolation (partial $R^2 = .005$ for slot given letter, $.008$ for
letter given slot), with a residual anti-``E'' letter preference ($-7.3$
pp, $P=.03$) (Table~\ref{tab:letter}).

\safeinput{tables/tab_letter.tex}{letter-arm decomposition}

\subsection{Fee-equivalents}
Converting effects through the fee gradient (monotonic, as required by the
prespecified gate) prices each signal in dollars per visit
(Table~\ref{tab:tradeoffs}). The 3.9$\to$4.7 rating step is worth \$157 per
visit (CI 147--167); recency \$19; telehealth \$14; responding to feedback
\$10. The demographic tilts are worth real money: a female-signaled name is
equivalent to a \$12.50 fee discount (CI 9--16), Hispanic- and
South-Asian--signaled names to \$14, and being listed first to \$11 per
visit, a pure interface artifact priced like a clinical credential.

\safeinput{tables/tab_tradeoffs.tex}{fee-equivalents table}

\subsection{Persona and interaction effects}
\begin{figure}[t]
\centering
\includegraphics[width=0.85\linewidth]{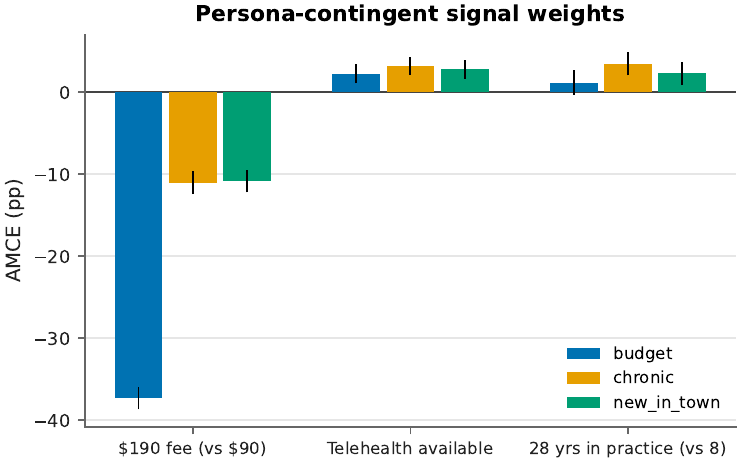}
\caption{Fee AMCEs by patient persona. The uninsured out-of-pocket persona
shows the steepest price penalty.}
\label{fig:persona}
\end{figure}

\begin{figure}[t]
\centering
\includegraphics[width=0.85\linewidth]{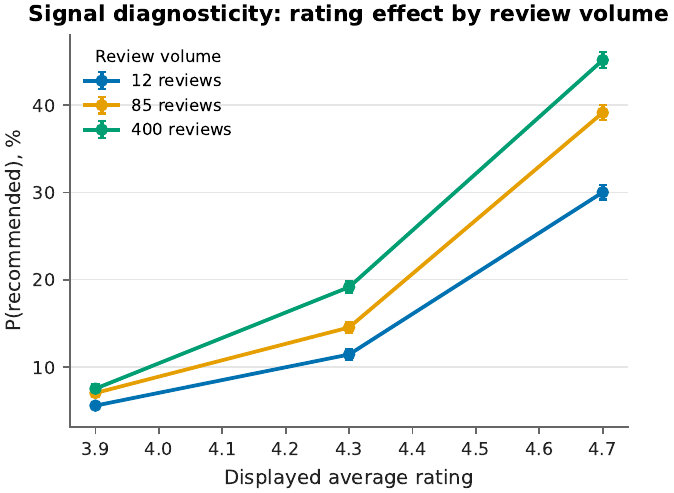}
\caption{Rating effects conditional on review volume: high volume amplifies
the high-rating premium.}
\label{fig:ratingvolume}
\end{figure}

Fee sensitivity is persona-dependent (H11, $P_{\text{Holm}}<.001$): the uninsured persona's price penalty far exceeds the insured
personas' (Figure~\ref{fig:persona}). Rating and volume interact (H12,
$P_{\text{Holm}}<.001$), with high volume amplifying the high-rating
premium (Figure~\ref{fig:ratingvolume}); the sign replicates in the
conditional logit, though probability-scale interactions under a
logit-additive process must be interpreted cautiously (Methods).

\subsection{Stated versus revealed importance}
\begin{figure}[t]
\centering
\includegraphics[width=\linewidth]{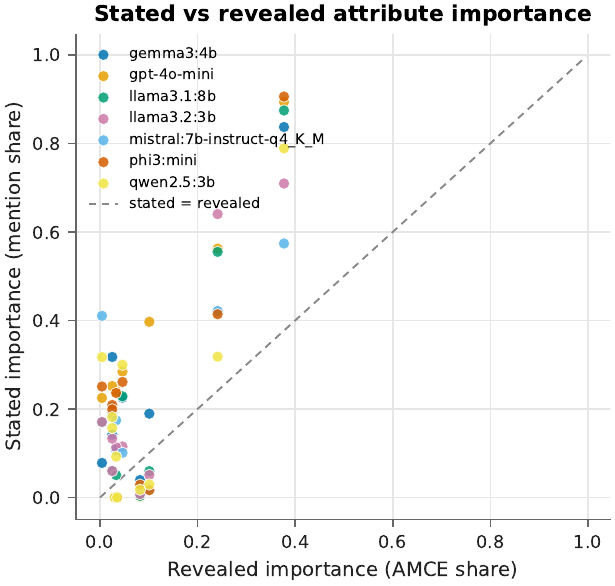}
\caption{Stated versus revealed importance by attribute and model: share of
stated reasons mentioning each attribute (stated) against its share of
total AMCE range (revealed). Demographic attributes and position sit far
below the diagonal.}
\label{fig:stated}
\end{figure}

Stated reasons track reputation but conceal demographics and position
(Figure~\ref{fig:stated}; 21{,}143 coded reasons). Rating is mentioned in 57--91\% of
reasons across models and price in 42--56\%, roughly commensurate with
their revealed weights. Gender is mentioned in $\le 0.03\%$ of reasons and
ethnicity in $\le 0.03\%$, against revealed importance shares of 3.0\% and
3.5\%, and position in $\le 4\%$ against a revealed share of 8.2\%. An
explanation mandate that relied on model self-report would have detected
none of the demographic or position effects measured here.

\subsection{Robustness}
\begin{figure}[t]
\centering
\includegraphics[width=\linewidth]{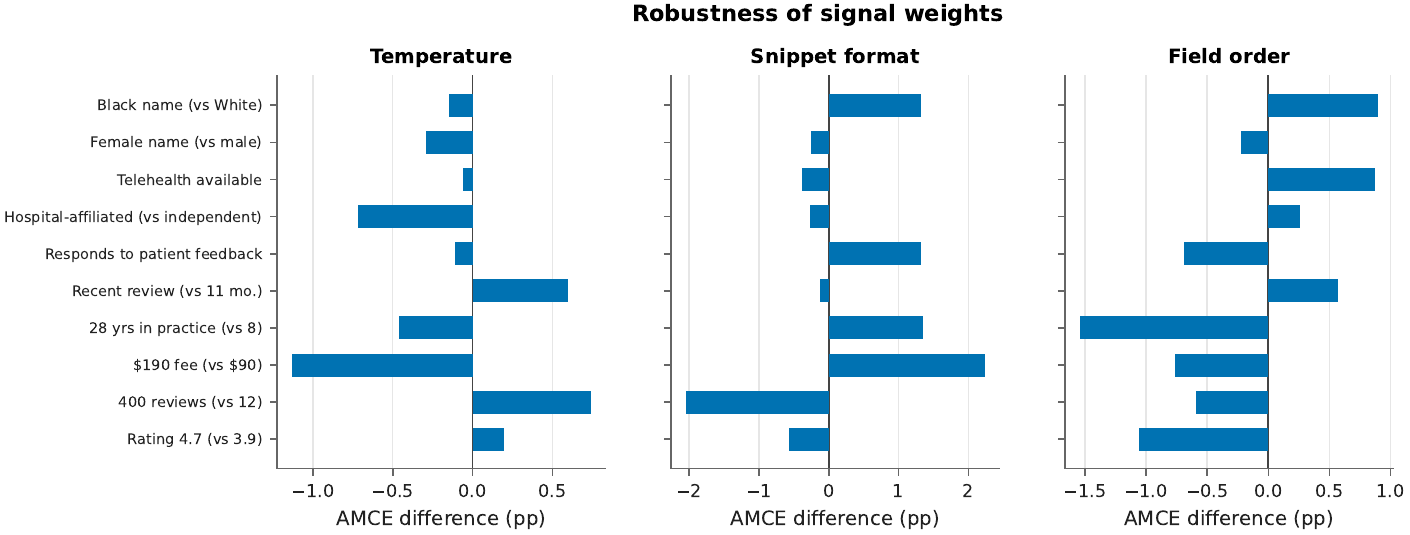}
\caption{Headline AMCEs across robustness arms (grounded snippets, reversed
layout, field order, temperature 0, persona-free, plausibility-filtered):
signs and ordering are stable throughout.}
\label{fig:robustness}
\end{figure}

\begin{sloppypar}
The reputation hierarchy is stable across every prespecified perturbation
(Figure~\ref{fig:robustness}): grounded web-snippet formatting, reversed
card layout, field order, and the plausibility-filtered subsample each
shift headline AMCEs by at most a few points without sign changes, and
temperature-0 decoding reproduces the ordering.
Removing the persona raises rating weight ($+8.4$ pp on the 4.7 step) and
halves the fee penalty ($-9.7$ pp), consistent with personas carrying real
budget information. The missingness model finds parse failures unrelated to
card content. The gate-excluded deepseek-r1 and the mistral retest arm
are the only material data-quality caveats.
\end{sloppypar}

\subsection{Exploratory: frontier-model session pilot}
To extend the audit's reach beyond open-weight models, four frontier
Claude-family models were piloted on the first 50 choice sets of the same
frozen design, and quantitative forecasts of their full-run behavior were
archived before any full-scale measurement (Methods). All four models
completed the pilot with no parse failures (200/200 calls; collection
caveats in Methods; numbers from
\texttt{results/tables/pilot\_frontier.csv}). Descriptively, the pilot
reproduces the reputation-signal ordering: cards rated 4.7 were chosen
39--43\% of the time against a 20\% null while cards rated 3.9 were chosen
0--1\%, the \$90 fee roughly quintupled choice share relative to \$190
(35--37\% vs.\ 5--8\%), and higher review volume and recency moved choice in
the expected directions in all four models
(Table~\ref{tab:pilot_frontier}). Two demographic patterns appear
consistently across the piloted models and merit confirmatory scrutiny in a
protocol-compliant run: physicians with female-signaled names were chosen
more often than male-signaled ones (22--24\% vs.\ 16--19\%), directionally
matching the confirmatory panel's pro-female tilt, and
East-Asian--signaled names were chosen more often (31--36\%) than
White-signaled (12--19\%) or Black-signaled (9--13\%) names, a pattern the
confirmatory panel does not show. At $n=50$ choice sets per model these are
directional readings only; none is a hypothesis test.

\input{tables/tab_pilot_frontier.tex}

The archived pilot-based forecasts of full-run choice probabilities
(Methods) could not be graded at analysis freeze because no
protocol-compliant API run of a forecast-target model was available; they
remain archived, timestamped, and ungraded, available for scoring the day
such a run is executed.

%% file: tables/tab_refusals.tex
\begin{table}[t]
\centering
\caption{Abstention and demographic-aware response rates.}
\label{tab:refusals}
\begin{tabular}{lrrr}
\toprule
Model & $n$ & Abstain (\%) & Demographic-aware (\%) \\
\midrule
gemma3:4b & 3024 & 0.0 & 0.0 \\
gpt-4o-mini & 3024 & 0.0 & 0.0 \\
llama3.1:8b & 3024 & 0.1 & 0.0 \\
llama3.2:3b & 3024 & 0.0 & 0.0 \\
mistral:7b-instruct-q4\_K\_M & 3024 & 0.6 & 0.0 \\
phi3:mini & 3024 & 1.2 & 0.1 \\
qwen2.5:3b & 3024 & 0.2 & 0.0 \\
\bottomrule
\end{tabular}
\\[2pt]
{\footnotesize\textit{Note.} Main arm. \emph{Abstain} = declines to choose or calls the options equal; \emph{demographic-aware} = the response text refers to a physician's name, gender, ethnicity, or race.}
\end{table}

%% file: tables/tab_pilot_frontier.tex
\begin{table}[t]
\centering
\caption{Exploratory session pilot of four Claude-family models: share of cards chosen by attribute level (null $=0.20$). Each model saw the first 50 main-arm choice sets of the frozen design (250 cards; 0 parse failures). Responses were collected interactively inside an agentic coding harness with default decoding parameters, not via the Messages API under the audit protocol; estimates are descriptive and exploratory, are quarantined from the confirmatory pipeline, and enter no confirmatory test.}
\label{tab:pilot_frontier}
\begin{tabular}{lrrrr}
\toprule
 & Haiku 4.5 & Sonnet 5 & Opus 4.8 & Fable 5 \\
\midrule
Rating 3.9 & 0.01 & 0.00 & 0.01 & 0.01 \\
Rating 4.3 & 0.14 & 0.18 & 0.18 & 0.16 \\
Rating 4.7 & 0.43 & 0.40 & 0.39 & 0.40 \\
\addlinespace
Fee \$90 & 0.36 & 0.37 & 0.35 & 0.35 \\
Fee \$140 & 0.17 & 0.17 & 0.17 & 0.17 \\
Fee \$190 & 0.06 & 0.05 & 0.08 & 0.08 \\
\addlinespace
12 reviews & 0.11 & 0.09 & 0.07 & 0.05 \\
85 reviews & 0.18 & 0.20 & 0.18 & 0.23 \\
400 reviews & 0.30 & 0.29 & 0.34 & 0.30 \\
\addlinespace
Recent review (3 days) & 0.27 & 0.26 & 0.26 & 0.27 \\
Stale review (11 mo.) & 0.14 & 0.15 & 0.15 & 0.14 \\
\addlinespace
Female name & 0.22 & 0.22 & 0.23 & 0.24 \\
Male name & 0.19 & 0.19 & 0.17 & 0.16 \\
\addlinespace
White name & 0.19 & 0.12 & 0.12 & 0.12 \\
Black name & 0.09 & 0.11 & 0.13 & 0.13 \\
Hispanic name & 0.16 & 0.14 & 0.16 & 0.16 \\
East Asian name & 0.31 & 0.36 & 0.31 & 0.33 \\
South Asian name & 0.24 & 0.25 & 0.25 & 0.24 \\
\bottomrule
\end{tabular}
\end{table}

%% file: sections/discussion_skeleton.tex
\subsection{Principal findings}
Three results anchor the audit. First, LLM physician recommendation is
reputation-driven to a degree exceeding human benchmarks: rating carries
37.7\% of attribute importance and fee 24.1\%, and the models weight the
rating step at \$157 per visit, signal weights directionally consistent
with, but steeper than, those estimated from human physician-choice
conjoints \citep{yaraghi2018}. Second, demographic parity is rejected in
the direction opposite to the discrimination documented in human audit
studies: female-, Hispanic-, South-Asian--, and Black-signaled names gain
one to three percentage points over male White-signaled names (worth
\$7--\$14 per visit), and no gender$\times$ethnicity cell meets the
prespecified $\pm$1.5 pp equivalence bound. Third, a content-free position
effect persists (first-listed worth \$11 per visit), and neither the
demographic nor the position effects ever surface in the models' stated
reasons.

\subsection{Comparison with prior work}
Where human patients discount female and minority physicians in
correspondence and field settings \citep{bertrand2004, gaddis2017}, the
audited models tilt modestly the other way, a pattern more consistent with
post-training fairness interventions overshooting neutrality than with
learned representativeness. Relative to clinical-bias audits
\citep{zack2024, omiye2023}, which find LLMs importing human clinical
disparities, the consumer-facing recommendation layer studied here behaves
differently: the risk is not replicated prejudice but \emph{unaudited,
opaque tilts of either sign}. Relative to descriptive chatbot-referral
audits \citep{oca2023}, the randomized design shows what they cannot:
that name effects exist net of every other attribute. And relative to the
travel-domain companion audit \citep{baig2026}, the same valence-price
primacy and content-free position bias reappear with identical anchor
levels, establishing cross-domain stability of the infomediary signal
hierarchy.

\subsection{An audit framework for AI infomediaries}
Beyond the physician-choice findings, the study demonstrates a reusable
template for auditing \emph{infomediaries}, that is, AI systems that
intermediate a person's choice among options or among other people. The same frozen-design,
randomized-conjoint machinery produced causal signal weights in two domains
(hotels \citep{baig2026}; physicians here) with identical anchor levels,
and the full pipeline spans design generation, execution, estimation,
equivalence testing, and monetary equivalents: auditing a new model against
the frozen design is a single command. Because model updates can
silently reweight signals, parity verdicts expire; we propose treating
audits of this form as \emph{recurring} monitoring rather than one-shot
evaluation. The observed heterogeneity underscores this: the pro-female
tilt ranges from near-zero (gemma3) to clearly positive across models, one
candidate model failed the auditability gate entirely, and the exploratory
frontier pilot hints at ethnicity orderings the confirmatory panel does not
show. Verdicts are therefore model-specific and version-specific, motivating
per-release audits.

\subsection{Implications}
For platforms: provider-directory orderings feed AI assistants; a
content-free first-position effect worth \$11 per visit means interface
choices allocate patient flow. For health-equity monitoring: group-level
parity testing with prespecified equivalence bounds detected tilts that
favor minority physicians, a reminder that ``bias'' in deployed systems
need not match the direction of historical discrimination, and that
monitoring must test for departures in both directions. For physicians and
practices: the fee-equivalent table prices controllable signals such as
recent reviews (\$19), telehealth (\$14), and responding to feedback
(\$10), against the dominant rating step (\$157). For regulators: the stated-vs-revealed
gap is the paper's sharpest governance result. Demographic attributes moved
choices but were mentioned in $\le$0.03\% of reasons; transparency
obligations that rely on model self-report would not have detected the
effects measured here. For the alignment debate on refusals: abstention was
rare (0.39\%) and indiscriminate rather than concentrated on
demographically contrastive choice sets: these models did not ``know when
not to answer.''

\subsection{Limitations}
Synthetic profiles bound external validity: real directories embed correlated
attributes and photographs; we audit text-only cards. Name signaling
identifies perceived-category effects, not mechanisms, and the significant
name-exemplar placebo shows specific names carry information beyond their
gender$\times$ethnicity cell, so cell-level estimates are audit-level
signals rather than clean category parameters. The audited panel comprises
six small open-weight models and one proprietary production model
(gpt-4o-mini); open-weight models under-represent the proprietary
assistants patients most use, and findings are a snapshot of specific model
versions (identifiers and dates recorded). The frontier-model
session pilot narrows but does not resolve this gap: four Claude-family
models were probed on the identical frozen design, and quantitative
forecasts of their full-run behavior were archived in advance of
measurement, a falsifiable-prediction step we propose as standard practice
for staged audits. The pilot remains exploratory by construction
(harness-embedded collection, default decoding, $n=50$ choice sets per
model), and the forecasts carry evidential weight only once graded against
a protocol-compliant API run. Choice sets held
specialty, board certification, and distance constant; effects of those
attributes are outside this design. Equivalence bounds of $\pm$1.5 pp are a
prespecified judgment call; smaller systematic effects could persist below
them.

\subsection{Conclusions}
A prespecified randomized audit of seven LLMs establishes causally that AI
physician recommendation runs on ratings and price, carries a content-free
position premium, and applies small systematic demographic tilts, favoring
female- and minority-signaled names, that the models never disclose in
their own explanations. None of these properties is observable from model
self-report; all of them are measurable, cheaply and repeatably, with a
frozen randomized instrument. As AI infomediaries absorb the referral
function, recurring behavioral audits of this form, rather than explanation
mandates, are the monitoring technology fit for purpose, and a frozen-design
instrument of this kind makes each new model release auditable in a day.

%% file: main.bbl
\begin{thebibliography}{17}
\providecommand{\natexlab}[1]{#1}
\providecommand{\url}[1]{\texttt{#1}}
\expandafter\ifx\csname urlstyle\endcsname\relax
  \providecommand{\doi}[1]{doi: #1}\else
  \providecommand{\doi}{doi: \begingroup \urlstyle{rm}\Url}\fi

\bibitem[Aggarwal et~al.(2024)Aggarwal, Murahari, Rajpurohit, Kalyan,
  Narasimhan, and Deshpande]{aggarwal2024}
Pranjal Aggarwal, Vishvak Murahari, Tanmay Rajpurohit, Ashwin Kalyan, Karthik
  Narasimhan, and Ameet Deshpande.
\newblock {GEO}: Generative engine optimization.
\newblock In \emph{Proceedings of the 30th ACM SIGKDD Conference on Knowledge
  Discovery and Data Mining}, pages 5--16, 2024.
\newblock \doi{10.1145/3637528.3671900}.

\bibitem[Ayers et~al.(2023)Ayers, Poliak, Dredze, Leas, Zhu, Kelley, Faix,
  Goodman, Longhurst, Hogarth, and Smith]{ayers2023}
John~W. Ayers, Adam Poliak, Mark Dredze, Eric~C. Leas, Zechariah Zhu,
  Jessica~B. Kelley, Dennis~J. Faix, Aaron~M. Goodman, Christopher~A.
  Longhurst, Michael Hogarth, and Davey~M. Smith.
\newblock Comparing physician and artificial intelligence chatbot responses to
  patient questions posted to a public social media forum.
\newblock \emph{JAMA Internal Medicine}, 183\penalty0 (6):\penalty0 589--596,
  2023.

\bibitem[Baig et~al.(2026)Baig, Gillani, and Ali]{baig2026}
Mirza Samad~Ahmed Baig, Syeda~Anshrah Gillani, and Asher Ali.
\newblock Whose hotel does the {AI} recommend? {An} algorithm audit of
  reputation signals in {LLM}-assisted hotel selection.
\newblock Working paper, under review, 2026.

\bibitem[Bertrand and Mullainathan(2004)]{bertrand2004}
Marianne Bertrand and Sendhil Mullainathan.
\newblock Are {Emily} and {Greg} more employable than {Lakisha} and {Jamal}?
  {A} field experiment on labor market discrimination.
\newblock \emph{American Economic Review}, 94\penalty0 (4):\penalty0 991--1013,
  2004.

\bibitem[Edelman et~al.(2017)Edelman, Luca, and Svirsky]{edelman2017}
Benjamin Edelman, Michael Luca, and Dan Svirsky.
\newblock Racial discrimination in the sharing economy: {Evidence} from a field
  experiment.
\newblock \emph{American Economic Journal: Applied Economics}, 9\penalty0
  (2):\penalty0 1--22, 2017.

\bibitem[Emmert et~al.(2013)Emmert, Sander, and Pisch]{emmert2013}
Martin Emmert, Uwe Sander, and Frank Pisch.
\newblock Eight questions about physician-rating websites: {A} systematic
  review.
\newblock \emph{Journal of Medical Internet Research}, 15\penalty0
  (2):\penalty0 e24, 2013.

\bibitem[Gaddis(2017)]{gaddis2017}
S.~Michael Gaddis.
\newblock How black are {Lakisha} and {Jamal}? {Racial} perceptions from names
  used in correspondence audit studies.
\newblock \emph{Sociological Science}, 4:\penalty0 469--489, 2017.

\bibitem[Gao et~al.(2012)Gao, McCullough, Agarwal, and Jha]{gao2012}
Guodong~Gordon Gao, Jeffrey~S. McCullough, Ritu Agarwal, and Ashish~K. Jha.
\newblock A changing landscape of physician quality reporting: {Analysis} of
  patients' online ratings of their physicians over a 5-year period.
\newblock \emph{Journal of Medical Internet Research}, 14\penalty0
  (1):\penalty0 e38, 2012.

\bibitem[Hainmueller et~al.(2014)Hainmueller, Hopkins, and
  Yamamoto]{hainmueller2014}
Jens Hainmueller, Daniel~J. Hopkins, and Teppei Yamamoto.
\newblock Causal inference in conjoint analysis: {Understanding}
  multidimensional choices via stated preference experiments.
\newblock \emph{Political Analysis}, 22\penalty0 (1):\penalty0 1--30, 2014.

\bibitem[Hanauer et~al.(2014)Hanauer, Zheng, Singer, Gebremariam, and
  Davis]{hanauer2014}
David~A. Hanauer, Kai Zheng, Dianne~C. Singer, Achamyeleh Gebremariam, and
  Matthew~M. Davis.
\newblock Public awareness, perception, and use of online physician rating
  sites.
\newblock \emph{JAMA}, 311\penalty0 (7):\penalty0 734--735, 2014.

\bibitem[Metaxa et~al.(2021)Metaxa, Park, Robertson, Karahalios, Wilson,
  Hancock, and Sandvig]{metaxa2021}
Dana{\"e} Metaxa, Joon~Sung Park, Ronald~E. Robertson, Karrie Karahalios,
  Christo Wilson, Jeff Hancock, and Christian Sandvig.
\newblock Auditing algorithms: {Understanding} algorithmic systems from the
  outside in.
\newblock \emph{Foundations and Trends in Human--Computer Interaction},
  14\penalty0 (4):\penalty0 272--344, 2021.

\bibitem[Obermeyer et~al.(2019)Obermeyer, Powers, Vogeli, and
  Mullainathan]{obermeyer2019}
Ziad Obermeyer, Brian Powers, Christine Vogeli, and Sendhil Mullainathan.
\newblock Dissecting racial bias in an algorithm used to manage the health of
  populations.
\newblock \emph{Science}, 366\penalty0 (6464):\penalty0 447--453, 2019.

\bibitem[Omiye et~al.(2023)Omiye, Lester, Spichak, Rotemberg, and
  Daneshjou]{omiye2023}
Jesutofunmi~A. Omiye, Jenna~C. Lester, Simon Spichak, Veronica Rotemberg, and
  Roxana Daneshjou.
\newblock Large language models propagate race-based medicine.
\newblock \emph{npj Digital Medicine}, 6:\penalty0 195, 2023.

\bibitem[Parikh et~al.(2024)Parikh, Oca, Conger, McCoy, Chang, and
  Zhang-Nunes]{oca2023}
Alomi~O. Parikh, Michael~C. Oca, Jordan~R. Conger, Allison McCoy, Jessica
  Chang, and Sandy Zhang-Nunes.
\newblock Accuracy and bias in artificial intelligence chatbot recommendations
  for oculoplastic surgeons.
\newblock \emph{Cureus}, 16\penalty0 (4):\penalty0 e57611, 2024.
\newblock \doi{10.7759/cureus.57611}.

\bibitem[Sandvig et~al.(2014)Sandvig, Hamilton, Karahalios, and
  Langbort]{sandvig2014}
Christian Sandvig, Kevin Hamilton, Karrie Karahalios, and Cedric Langbort.
\newblock Auditing algorithms: {Research} methods for detecting discrimination
  on internet platforms.
\newblock In \emph{Data and Discrimination: Converting Critical Concerns into
  Productive Inquiry (preconference of the 64th Annual Meeting of the
  International Communication Association)}. Seattle, WA, 2014.

\bibitem[Yaraghi et~al.(2018)Yaraghi, Wang, Gao, and Agarwal]{yaraghi2018}
Niam Yaraghi, Weiguang Wang, Guodong~Gordon Gao, and Ritu Agarwal.
\newblock How online quality ratings influence patients' choice of medical
  providers: {Controlled} experimental survey study.
\newblock \emph{Journal of Medical Internet Research}, 20\penalty0
  (3):\penalty0 e99, 2018.

\bibitem[Zack et~al.(2024)Zack, Lehman, Suzgun, Rodriguez, Celi, Gichoya,
  Jurafsky, Szolovits, Bates, Abdulnour, Butte, and Alsentzer]{zack2024}
Travis Zack, Eric Lehman, Mirac Suzgun, Jorge~A. Rodriguez, Leo~Anthony Celi,
  Judy Gichoya, Dan Jurafsky, Peter Szolovits, David~W. Bates, Raja-Elie~E.
  Abdulnour, Atul~J. Butte, and Emily Alsentzer.
\newblock Assessing the potential of {GPT-4} to perpetuate racial and gender
  biases in health care: a model evaluation study.
\newblock \emph{The Lancet Digital Health}, 6\penalty0 (1):\penalty0 e12--e22,
  2024.

\end{thebibliography}
